\documentclass{article}
\usepackage{spconf,amsmath,graphicx}
\usepackage{threeparttable}
\newcommand \footnoteONLYtext[1]
{
	\let \mybackup \thefootnote
	\let \thefootnote \relax
	\footnotetext{#1}
	\let \thefootnote \mybackup
	\let \mybackup \imareallyundefinedcommand
}
\makeatletter
\renewcommand{\maketag@@@}[1]{\hbox{\m@th\normalsize\normalfont#1}}%
\makeatother

\title{Semi-supervised Sound Event Detection with Local and Global Consistency Regularization}
\name{Yiming Li$^{1,2}$, 
      Xiangdong Wang$^{1,\star}$,
      Hong Liu$^{1}$, 
      Rui Tao$^{3}$, 
      Long Yan$^{3}$,
      Kazushige Ouchi$^{3}$
      }
\address{$^1$ Beijing Key Laboratory of Mobile Computing and Pervasive Device,\\
Institute of Computing Technology, Chinese Academy of Sciences, Beijing, China. \\     
$^2$ University of Chinese Academy of Sciences, Beijing, China. \\ 
$^3$ Toshiba China R\&D Center, Beijing, China. \\ }
\usepackage{enumitem}
\begin{document}
%
\maketitle
\begin{abstract}
Learning meaningful frame-wise features on a partially labeled dataset is crucial to semi-supervised sound event detection. Prior works either maintain consistency on frame-level predictions or seek feature-level similarity among neighboring frames, which cannot exploit the potential of unlabeled data. In this work, we design a Local and Global Consistency (LGC) regularization scheme to enhance the model on both label- and feature-level. The audio CutMix is introduced to change the contextual information of clips. Then, the local consistency is adopted to encourage the model to leverage local features for frame-level predictions, and the global consistency is applied to force features to align with global prototypes through a specially designed contrastive loss. Experiments on the DESED dataset indicate the superiority of LGC, surpassing its respective competitors largely with the same settings as the baseline system. Besides, combining LGC with existing methods can obtain further improvements. The code will be released soon.
\end{abstract}
\section{Introduction}
\label{sec:intro}
\footnoteONLYtext{$^\star \,\, \text{Corresponding author.}$}
Sound event detection (SED) aims at identifying the category of sound events as well as their temporal boundaries, which can assist in understanding acoustic scenes and perceiving physical environments \cite{mesaros2021sound}. Thanks to the advances in Deep Learning (DL) techniques, the capability of SED systems has been greatly improved \cite{li2023astsed,xu2023pretrain}. While applying DL to SED tasks, the frame-level features are first extracted by an encoder and then transformed into probability vectors by a classifier. As a result, the quality of frame-wise features matters a lot to the performance of SED systems. However, learning discriminative and robust features is a tricky problem as the SED dataset is usually partially labeled to avoid laborious annotating.   

To leverage the unlabeled SED dataset for feature learning, semi-supervised learning becomes a promising remedy, which usually adopts the teacher-student framework with the former holding an exponential moving average (EMA) of the latter. Hence, the teacher model is supposed to generate more reliable pseudo labels to guide the student model in learning informative features on unlabeled data, which can be viewed as a form of label-level consistency. Among methods applying label-level consistency, MeanTeacher \cite{tarvainen2017mean} utilizes a soft consistency loss to regularize the predictions of the teacher and the student model, while CMT \cite{xiao2023cmt} improves it by filtering out low-confident pseudo labels. Furthermore, ICT \cite{verma2019ict} incorporates mixup \cite{zhangmixup} to encourage the predictions for interpolated frames to be consistent with the interpolation of predictions for original frames. With the emergence of audio augmentations \cite{park19e_interspeech, nam2022filteraugment}, many methods resort to smoothness assumptions \cite{chapelle2009semi} which require similar predictions for the same data under different perturbations. They usually warp the spectrograms and maintain consistency between predictions of original inputs and perturbed inputs to ensure the robustness of learned features. For instance, SCT \cite{koh2021sct} retains consistency with shift-related operations, including time and frequency shift, while RCT \cite{shao22_interspeech} seeks a feasible combination of random augmentations, such as pitch shift and time mask, and applies a self-consistency loss. Although methods like RCT can achieve supreme performance, the training cost is rather noticeable. 

In addition to label-level regularization, TCL \cite{kothinti2022tcl} and MPR \cite{park22cmpr} encourage the neighboring frame-wise representations to be similar if the two share the same strong labels while being different if there is an event boundary, which can be considered as a kind of feature-level consistency. Although such methods can be combined with label-level consistency and achieve additional performance gains, they solely regularize the neighboring embeddings and are only applicable to strongly labeled settings since ground truth is requested to compute the related loss, which precludes their applications.

\begin{figure}[htp]
  \centering
  \centerline{\includegraphics[scale=0.4]{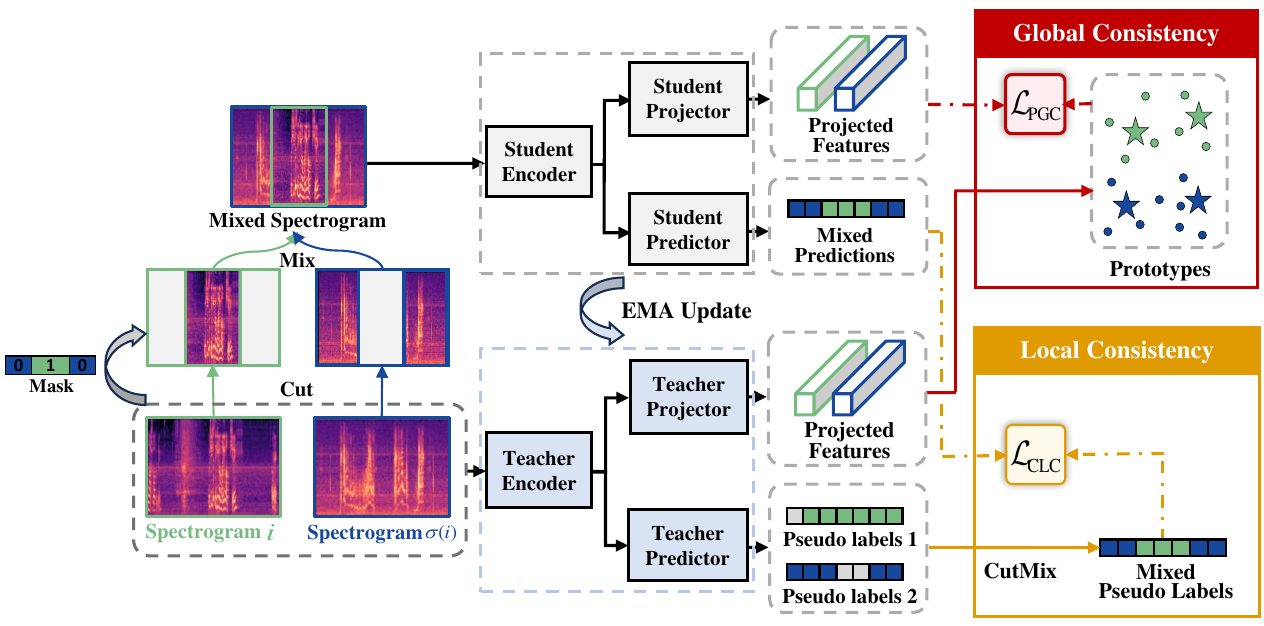}}
  \vspace{-0.75em}
  \caption{Illustration of proposed LGC method, where the pipeline of traditional MeanTeacher method is omitted for simplicity.}
  \label{fig:lgcr_flowchart}
\end{figure}

In this paper, we propose to explore both label- and feature-level consistency to regularize the feature learning process in a collaborative way. Specifically, audio CutMix is adopted to modify the boundaries or contextual information of sound events. Based on the CutMixed inputs, label-level \emph{local consistency} is exploited to learn robust frame features with limited contexts, and feature-level \emph{global consistency} is devised to reduce the intra-class variance while increasing the inter-class variance of frame features in a global view. We refer to our method as Local and Global Consistency ({\bf LGC}) regularization, and an overview is given in Figure~\ref{fig:lgcr_flowchart}:
\begin{itemize}[labelsep = .5em, leftmargin = 0pt, itemindent = 1em]
\item {\bf Local Consistency}\quad As shown in Figure~\ref{fig:lgcr_flowchart}, we adapt CutMix \cite{yun2019cutmix} to cut complementary segments from two input spectrograms before
jointing them accordingly and force the predictions of mixed inputs to be consistent with the mixed pseudo labels by the local consistency loss $\mathcal{L}_{\text {CLC}}$. This consistency term helps detach temporal dependency in mixed inputs so that the SED model can focus on local semantics, thereby learning robust patterns under varying contexts. Compared to SCT \cite{koh2021sct} and RCT \cite{shao22_interspeech}, our method does not rely on audio warping to reach label-level consistency, making it more efficient and scalable to complement other methods. 

\item {\bf Global Consistency}\quad We adopt multiple prototypes, which can be viewed as the global representatives of class-wise frame features across the whole dataset, to model the feature space. Under the supervision of a global consistency loss $\mathcal{L}_{\text {PGC}}$ shown in Figure~\ref{fig:lgcr_flowchart}, frame features from the student model are appealed to corresponding class prototypes and repelled to prototypes of other classes. As a result, per-class frame features can be clustered compactly on the feature space, making it easier for the classifier to learn a decision boundary at the low-density region \cite{chapelle2005semi}, which can also help promote the label-level consistency. Different from TCL \cite{kothinti2022tcl} and MPR \cite{park22cmpr}, our approach does not require frames to be neighbors or manually labeled, indicating that each frame can be regularized on feature-level.
\end{itemize}

Extensive experiments suggest that LGC achieves superior results on the DESED validation dataset. While combining it with audio augmentations or other consistency regularization methods, the performance can also be significantly improved. 
 \vspace{-0.5em}
\vspace{-1.75em}
\section{Methodology}
\label{sec:method}
\subsection{Preliminary}
Currently, a typical semi-supervised SED system usually consists of a feature encoder $f^{(l)}(\cdot)$ and a predictor
$g^{(l)}(\cdot)$, where $l \in \{\mathrm{S}, \mathrm{T}\}$ represents the student and 
teacher models respectively. We refer to the
input as $\mathbf{X} \in \mathbb{R}^{T\times
F}$, where $T$ and $F$ denote the time and 
frequency axis dimensions of the spectrogram. Given $\mathbf{X}$, $f^{(l)}$ extracts the embedded features by $f^{(l)}(\mathbf{X}):\mathbf{X} \rightarrow \mathbf{Q}^{(l)} \in \mathbb{R}^{T'\times D}$, after which $g^{(l)}$ transforms $\mathbf{Q}^{(l)}$ to the frame-level predictions by $g^{(l)}\circ f^{(l)}(\mathbf{X}): \mathbf{Q}^{(l)} \rightarrow \mathbf{P}^{(l)} \in \mathbb{R}^{T' \times M}$, where $T'$ represents the final number of frames (Note that $T' \textless T$ as the desired time resolution for SED is lower than that of original clips) and $D, M$ denote the number of feature dimensions and sound classes. In SED systems,  $f^{(l)}$ can be a CRNN \cite{cakir2017crnn} or Transformer \cite{hu22d_MGA} and $g^{(l)}$ is a dense layer with sigmoid activation. The loss function is defined as $\mathcal{L} = \mathcal{L}_{\text{Sup}} + r(step) \mathcal {L}_{\text{MT}}$, where $\mathcal{L}_{\text{Sup}}$ is the BCE loss for labeled data and $\mathcal{L}_{\text{MT}}$ is the L2 consistency loss between student and teacher predictions whose weight is adjusted along training steps by $r(step)$. In this work, we reserve the above framework while adding two loss terms to explore the label-level local consistency and the feature-level global consistency.
\vspace{-0.5em}
\subsection{Audio CutMix}
CutMix is a useful augmentation strategy in computer vision. It randomly crops a patch from one image and then pastes it to another to synthesize the augmented sample. Inspired by that, we employ CutMix to the audio domain but only crop spectrograms along the time axis to preserve the information in frequency domains as shown in Figure~\ref{fig:lgcr_flowchart}. At first, we retain a copy of the data batch and randomly shuffle it. As a result, each spectrogram $\mathbf{X}_{i}$ can be matched to another one $\mathbf{X}_{\sigma(i)}$ in the copied batch at index $i$. The corresponding frame-level predictions from the teacher models are $\mathbf{P}_i^{(\mathrm{T})}$ and $\mathbf{P}^{(\mathrm{T})}_{\sigma(i)}$. We define the CutMix operation \text{CM} as:
\vspace{-0.5em}
\begin{equation}
  \text{CM}(\mathbf{X}_i, \mathbf{X}_{\sigma(i)}) = \boldsymbol{m}\odot \mathbf{X}_i + (\boldsymbol{1} - \boldsymbol{m})\odot \mathbf{X}_{\sigma(i)}
\end{equation}
\begin{equation}
  \text{CM}(\mathbf{P}_i^{(\mathrm{T})}, \mathbf{P}^{(\mathrm{T})}_{\sigma(i)}) = \boldsymbol{m}'\odot \mathbf{P}_i^{(\mathrm{T})} + (\boldsymbol{1} - \boldsymbol{m}')\odot \mathbf{P}^{(\mathrm{T})}_{\sigma(i)}
\end{equation}
where $\boldsymbol{m} \in \{0, 1\}^T$ is a binary mask of length $T$, $\boldsymbol{m}'$ is the pooling version of $\boldsymbol{m}$ with length $T'$ and $\odot$ is the element-wise product. We initialize $\boldsymbol{m}$ as a zero vector and randomly replace a consecutive part with 1. By CutMix, we change the boundaries or contextual information of a sound event instead of warping its features. 
\vspace{-0.5em}
\subsection{Label-level Local Consistency}
To encourage the model to alleviate excessive temporal dependency and make it less vulnerable to varying contexts while making predictions, we introduce the label-level consistency loss $\mathcal{L}_{\text{CLC}}$ based on the audio CutMix operation. Specifically, it forces the student's predictions of CutMixed spectrograms to be consistent with the CutMixed teacher's predictions of original samples, which is written as:
\begin{equation}
\mathcal{L}_{\text{CLC}}=\sum_i \Vert g^{(\mathrm{S})}\circ f^{(\mathrm{S})}\left(\text{CM}\left(\mathbf{X}_i, \mathbf{X}_{\sigma(i)}\right)\right)-\text{CM}\left(\mathbf{P}_i^{(\mathrm{T})}, \mathbf{P}^{(\mathrm{T})}_{\sigma(i)}\right)\Vert_2^2
\end{equation}
The above objective is challenging for the SED model because it is expected to react accurately to the confusing event boundaries introduced by CutMix, even with limited contextual information, which may enhance its localization sensitivity.
Similarly, $r(step)$ is used to control the weight of $\mathcal{L}_{\text{CLC}}$, and the overall loss can be written as $\mathcal{L}_1 = \mathcal{L}_{\text{Sup}} + r(step) (\mathcal {L}_{\text{MT}}+\mathcal {L}_{\text{CLC}})$.
\vspace{-0.5em}
\subsection{Feature-level Global Consistency}
Prototypical network \cite{snell2017prototypical} has been widely adopted in few-shot learning. It uses the global average representation of a class as the prototype and assigns labels based on the distance between sample embeddings and class prototypes. The intuition is that samples belonging to the same class are similar in feature space. Built on this idea, we propose the concept of global consistency, which regularizes the frame feature of SED models to be close to its class prototype and far away from prototypes of other classes. To account for the intra-class diversity, we utilize \emph{Multiple Prototypes} (\emph{MP}) to represent a class. Besides, we add a projector $h^{(l)}$ to project $\mathbf{Q}^{(l)}$ to a low dimension $D'$, and only the projected features are required to maintain global consistency to prevent the original features losing semantic information. In the following sections, we detail the process of estimating multiple prototypes for a given class and reaching global consistency with prototypical contrastive learning.
\vspace{-0.5em}
\subsubsection{Prototype Estimation}
\label{subsubsec:pe}
The success of prototype-based consistency relies on a set of informative prototypes derived from frame features. Thus, it is vital to design strategies to initialize and maintain the pool of prototypes.

\noindent{\bf Offline Prototype Initialization}\qquad We first train the SED model using $\mathcal{L}_1$ for several epochs to ensure it can extract meaningful features and grasp basic detection ability. Then, we suspend the training process before feeding the training set into the teacher model. As a result, for each frame $k$ in audio clips from the training set, its probability vector $\boldsymbol{p}_k^{(\mathrm{T})} \in \mathbb{R}^M$ and projected feature vector $\boldsymbol{v}_k^{(\mathrm{T})} \in \mathbb{R}^{D'}$ can be obtained through forward pass. For each class $i$, a set $Q_i$ is created to store high-quality projected features $\boldsymbol v_{k}^{(\mathrm{T})}$ whose corresponding $\boldsymbol {p}_{k, i}^{(\mathrm{T})}$ is larger than a threshold $\tau_+$. We refer to such features as high-quality features as they can be utilized to generate highly confident predictions. After collecting all the projected features of class $i$, the K-Means algorithm is applied to implement intra-class clustering on $Q_i$, resulting in $C$ cluster centroids $\boldsymbol c_{i,j},j = 1,\cdots, C$, which can be viewed as initial prototypes for class $i$.

\noindent{\bf Online Prototype Iteration}\qquad In the following training progress, prototypes are dynamically updated from the teacher's projected features extracted from both labeled and unlabeled frames to better capture the status of the SED model. Specifically, at each training step, $Q_i$ is emptied and used to re-collect high-quality features from the teacher model. To guarantee the stability of prototypes, for the $j$-th prototype of class $i$ , we update $\boldsymbol c_{i,j}$ in a moving average manner by
\vspace{-0.25em}
\begin{equation}
\boldsymbol{c}_{i,j} \leftarrow \text{Normalize}(\beta \boldsymbol c_{i,j} + (1 - \beta)\boldsymbol{\hat c}_{i,j})
\end{equation}
where $\beta$ is a momentum coefficient, $\text{Normalize}(\cdot)$ is the L2 normalization function and $\boldsymbol{\hat c}_{i,j}$ is the feature centroid of frames in $Q_i$ which satisfy that $\boldsymbol c_{i, j}$ should be the most similar prototype with these frames among all other prototypes of class $i$, its mathematical formulation is defined as:
\vspace{-0.25em}
\begin{equation}
\boldsymbol{\hat{c}}_{i, j}=\frac{\sum\limits_{k=1}^{\operatorname{len}\left(Q_i\right)} \boldsymbol{v}_k^{(\mathrm{T})} \mathbb{I}\left[\mathop{\mathrm{\arg\max}}\limits_{n=1, \ldots, C}  \langle\boldsymbol{c}_{i,n}, \boldsymbol{v}_k^{(\mathrm{T})}\rangle =j\right]}{\sum\limits_{k=1}^{\operatorname{len}\left(Q_i\right)} \mathbb{I}\left[\mathop{\mathrm{\arg\max}}\limits_{n=1, \ldots, C}  \langle\boldsymbol{c}_{i,n}, \boldsymbol{v}_k^{(\mathrm{T})} \rangle =j\right]}
\vspace{-0.75em}
\end{equation}
where $\boldsymbol{v}_k^{(\mathrm{T})}$ is the $k$-th element of $Q_i$,  $\operatorname{len}(Q_i)$ is the size of $Q_i$, $\langle \cdot,\cdot \rangle$ denotes the cosine similarity function and $\mathbb{I}[\cdot]$ denotes the indicator function which evaluates to 1 if $\cdot$ is true and 0 otherwise.
\vspace{-0.5em}
\subsubsection{Selective Prototypical Contrastive Learning}
\label{subsubsec:spcl}
Inspired by progress in self-supervised audio representation learning \cite{saeed2021contrastiveaudio, niizumi2022byola}, which applies unsupervised contrastive learning to force the clip embedding to be close to its augmented view while being dissimilar to embeddings of other clips, we design a class-aware prototypical contrastive learning scheme. It aims to learn a global feature space where frames from the same class are close to the class prototypes while frames from different classes are well separated.

Similar to Section 2.3, a batch of CutMixed inputs are fed into the student model to obtain the projected feature $\boldsymbol v_k^{(\mathrm{S})}$ of each frame $k$ as shown in Figure~\ref{fig:lgcr_flowchart}. We then generate the frame-to-class similarity score $s_{k, i} = \max\limits_{j=1,\cdots, C}\langle \boldsymbol {v}_k^{(\mathrm{S})},\boldsymbol {c}_{i, j}\rangle$ between the frame feature $\boldsymbol {v}_k^{(S)}$ and prototypes of class $i$, which uses the maximal similarity between a frame and prototypes of a class as the similarity between a frame and a class. Then, we optimize the following loss to reach our goal:
\vspace{-0.5em}
\begin{equation}
\mathcal{L}_{\text{PGC}} = -\sum\limits_{k}\sum_{i = 1}^{M} \mathbb{I}(\boldsymbol {p}_{k,i}^{(\mathrm{T})} >\tau_{+})\log\frac{e^{\frac{s_{k,i}}{\gamma}}}{\sum_{m=1}^{M}e^{\frac{s_{k,m}}{\gamma}}}
\vspace{-0.5em}
\end{equation}
where $\gamma$ is a scalar temperature parameter and the meanings of $\tau_{+}$ and $\boldsymbol {p}_{k, i}^{(\mathrm{T})}$ are consistent as mentioned in Section \ref{subsubsec:pe}. The above loss follows a similar formulation as the InfoNCE loss \cite{chen2020infonce} while novelly introducing an indicator function to fit for the semi-supervised setting. If the teacher probability $\boldsymbol {p}_{k,i}^{(\mathrm{T})}$ is larger than $\tau_{+}$, optimizing $\mathcal{L}_{\text {PGC}}$ will maximize $s_{k,i}$ but minimize $s_{k, m}$ ($ m = 1, \cdots, M$ and $m \neq i$), thereby pulling together $\boldsymbol v_{k}^{(\mathrm{S})}$ and class $i$'s prototype which is most similar to $\boldsymbol v_{k}^{(\mathrm{S})}$ while pushing away $\boldsymbol v_{k}^{(\mathrm{S})}$ from prototypes belonging to other classes in the feature space. Otherwise, the teacher model does not make a confident prediction, and the loss term will be evaluated to 0 for fear that the frame feature may be pushed to an improper prototype. Note that $\boldsymbol v_{k}^{(\mathrm{S})}$ can be pushed to more than one class prototype as long as it satisfies the indicator function, making it applicable to multi-label settings. Furthermore, the student's inputs are also CutMixed, making the above process more challenging as contextual reliance in audio clips is removed by CutMix. As a result, the SED model can learn how to use temporal cues more properly when extracting features. 

However, we find that not every frame needs to be involved in the contrastive process. We then devise the \emph{Selective Anchor Sampling} (\emph{SAS}) strategies to choose candidate frames:
\begin{itemize}[labelsep = .5em, leftmargin = 0pt, itemindent = 1em]
\item We only select frames from weakly labeled and unlabeled clips since the feature learning process of strongly labeled frames can be supervised by the ground truth well enough.
\item We only select frames where the student is likely to make wrong predictions due to the inferiority of frame features. Specifically, given a frame $k$, if the teacher model predicts it as class $i$ with a high confidence $\boldsymbol {p}_{k,i}^{(\mathrm{T})} > \tau_{+}$ but the student does not ($\boldsymbol {p}_{k,i}^{(\mathrm{S})} < \tau_{-}$ and $\tau_{-} \ll \tau_{+}$), then it will be involved.
\end{itemize}

By applying SAS, only about 2\% frames are required to compute $\mathcal{L}_{\text{PGC}}$, resulting in significant improvements in training efficiency and model performance as potential overcorrections of features are alleviated. The overall loss considering prototype-based global consistency is $\mathcal{L}_2 = \mathcal{L}_1 + \alpha \mathcal{L}_{\text{PGC}}$, where $\alpha$ is a trade-off parameter.

By incorporating $\mathcal{L}_{\text{CLC}}$ and $\mathcal{L}_{\text{PGC}}$ simultaneously, LGC exploits both label-level local consistency and feature-level global consistency, and we argue that the two consistency terms can benefit from each other. On the one hand, while pursuing global consistency, inputs of the student model are CutMixed, encouraging the encoder to refine feature extraction since much irrelevant noise is induced by CutMix. On the other hand, the class-wise frame features can be more compact and discriminative with global consistency regularization, enhancing the robustness of the classifier to recognize an event with limited contextual information to reach local consistency.
\vspace{-1.5em}
\section{Experiments}
The DESED dataset \cite{turpault2019desed} (the official dataset of DCASE 2022 Task 4) is chosen to conduct experiments, which contains 1578 weakly labeled clips, 10000 strongly labeled synthetic clips, and 14412 unlabeled clips. For each 10-second audio clip, we resample it to 16 kHz and transform it into 626 frames using the short-term Fourier transform with a window length of 2048 samples and a hop length of 256 samples. And 128 log mel-band magnitudes are then extracted as input features. We implement all methods based on the same CRNN network with the official baseline network of DCASE 2022 Task 4. Evaluation of each method is performed with Event-Based macro F1 (EB-F1) \cite{mesaros2016ebf1} and Polyphonic Sound Detection Scores (PSDSs) \cite{bilen2020psds} on the validation set composed of 1168 clips. For the proposed LGC, we empirically set $C=3$, $\beta=0.99$, $\tau_{+} = 0.9$, $\tau_{-} = 0.5$ and $\gamma = \alpha = 0.1$. The model is trained with $\mathcal{L}_1$ for the first 100 epochs and then with $\mathcal{L}_2$ for the last 100 epochs. As for other training settings, we follow those of the official baseline. 
\vspace{-0.75em}
\subsection{Comparison with Other Methods}

To verify the effectiveness of LGC, we first compare it with existing works that do not utilize audio warping either. Among them, TCL and MPR pursue feature-level consistency, while Baseline, CMT, and ICT are trained with label-level consistency regularization. The related results are shown in Table~\ref{tab:table_one}. It can be observed that LGC excels to the baseline significantly by 6.5\% on EB-F1, 0.045 and 0.034 on PSDSs. Moreover, it surpasses its counterparts to a large extent in terms of the time-sensitive metrics, namely PSDS$_{1}$ and EB-F1, indicating that LGC can boost the event localization capacity.

To exploit the potentials of LGC and seek a fair comparison with methods that leverage audio augmentations to maintain consistency, we promote LGC in two ways: (1)  we impose FilterAug \cite{nam2022filteraugment} on LGC without introducing any additional consistency; (2) we integrate LGC into SCT and RCT. The comparison results are reported in Table~\ref{tab:table_two}. As seen, simply adding augmentations to LGC (LGC + Aug) leads to remarkable improvements compared to the vanilla LGC, implying that LGC is also robust to audio perturbations. Moreover, LGC + Aug is also advanced compared with prior works. It outperforms SCT by approximately 0.04 on both PSDSs and achieves higher PSDS$_{1}$ and comparable PSDS$_{2}$ with much less training consumption than RCT. When combining SCT or RCT with LGC, further improvements can be obtained with little additional training cost, demonstrating its effectiveness while working with existing consistency regularization techniques. Finally, we extend the proposed LGC to FDY-CRNN \cite{nam22_fdc}, a powerful backbone adopted by many recent works, without additional training augmentations. The notable improvements in Table~\ref{tab:table_fdy} suggest its scalability.
\vspace{-1.2em}
\begin{table}[th]
\begin{threeparttable}   
  \caption{Comparing LGC with methods excluding audio warping.}
  \setlength{\tabcolsep}{14pt}
  \label{tab:table_one}
  \centering
  \begin{tabular}{cccc}
    \toprule
    \multicolumn{1}{c}{\textbf{Methods}} & {\textbf{EB-F1}(\%)} & {\textbf{PSDS$_{1}$}} & {\textbf{PSDS$_{2}$}} \\
    \midrule
    Baseline & 42.3 & 0.347 & 0.545             \\
    TCL* \cite{kothinti2022tcl} & 43.4 & 0.359 & 0.560  \\
    MPR* \cite{park22cmpr} & 43.6 & 0.365 & 0.570  \\
    ICT \cite{verma2019ict} & 44.2 & 0.368 & 0.567  \\
    CMT \cite{xiao2023cmt} & 44.4  & 0.371 & 0.572 \\
    LGC & \bf 48.8 & \bf 0.392 & \bf 0.579  \\
    \bottomrule
  \end{tabular}
  \begin{tablenotes} 
  \item {\footnotesize * are trained with BCE Loss instead of TBFL \cite{park2022tbfl} to guarantee fairness}.
  \end{tablenotes} 
  \end{threeparttable}    
\end{table}
\vspace{-2.5em}
\begin{table}[th]
  \caption{Comparing promoted LGC with methods involving audio warping augmentations. Runtime records training time per batch on a single NVIDIA RTX 3090 GPU.}
  \setlength{\tabcolsep}{5pt}
  \label{tab:table_two}
  \centering
  \begin{tabular}{ccccc}
    \toprule
    \multicolumn{1}{c}{\textbf{Methods}} & {\textbf{EB-F1}(\%)} & {\textbf{PSDS$_{1}$}} & {\textbf{PSDS$_{2}$}} & {\textbf{Runtime (sec.)}}\\
    \midrule
    SCT \cite{koh2021sct} & 44.6  & 0.368 & 0.571 & \bf 0.29 \\
    LGC + SCT & \bf 51.0 & 0.407 & 0.612 & 0.41  \\
    RCT \cite{shao22_interspeech} & 46.5 & 0.401 & 0.614 & 1.39  \\
    LGC + RCT & 49.4 & 0.408 & \bf 0.630 & 1.61  \\
    LGC + Aug & 50.7 & \bf 0.411 & 0.613 & 0.33             \\
    \bottomrule
  \end{tabular}
\end{table}
\vspace{-2.0em} 
\begin{table}[th]
  \centering
  \begin{threeparttable}   
  \caption{Extending the proposed LGC to FDY-CRNN backbone. }
  \setlength{\tabcolsep}{16pt}
  \label{tab:table_fdy}
  \centering
  \begin{tabular}{cccc}
    \toprule
    \multicolumn{1}{c}{\textbf{LGC}} & {\textbf{EB-F1}(\%)} & {\textbf{PSDS$_{1}$}} & {\textbf{PSDS$_{2}$}} \\
    \midrule
    - & 46.5 & 0.382 & 0.593             \\
    \checkmark & \bf 49.8 & \bf 0.424 & \bf 0.647  \\
    \bottomrule
  \end{tabular}
  \end{threeparttable}    
\end{table}
\vspace{-1.25em}
\subsection{Ablation Studies on Proposed Techniques}
\begin{figure}[htp]
  \centering
  \includegraphics[scale=0.275]{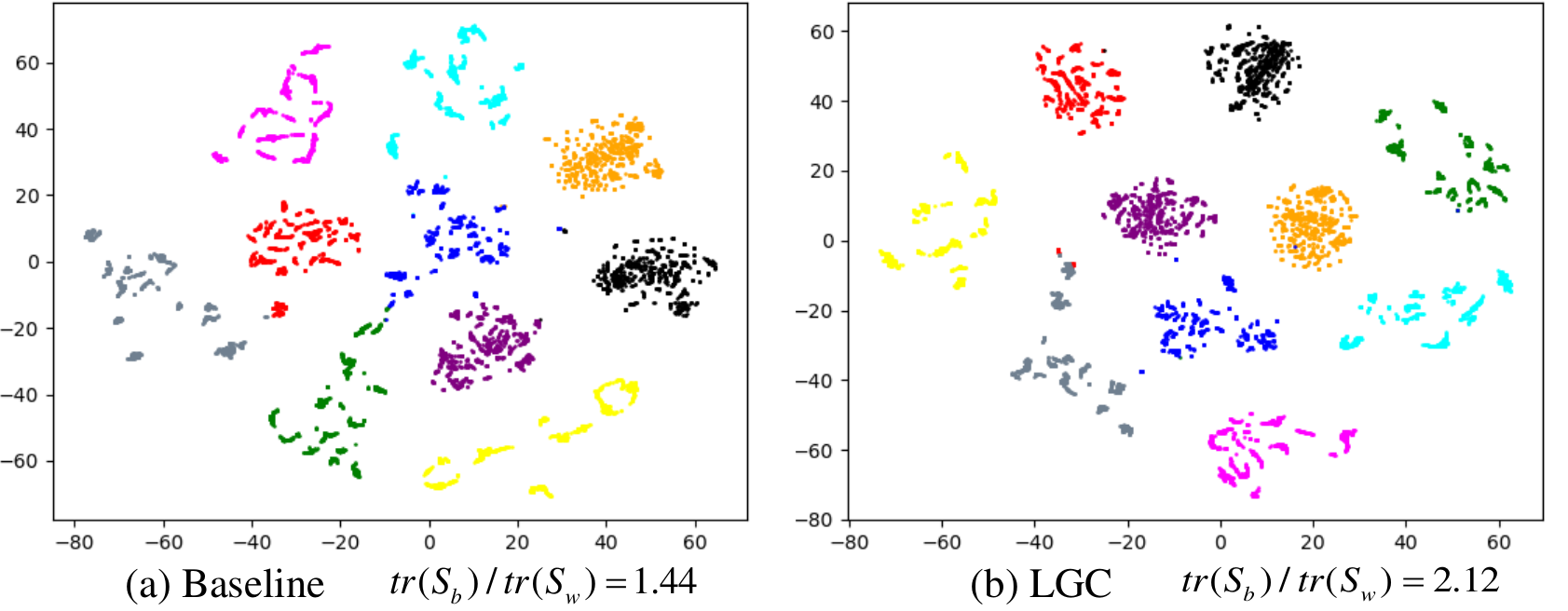}
    \vspace{-0.75em}
  \caption{Frame-wise feature embeddings visualization of (a) Baseline (b) LGC on the DESED validation set using t-SNE \cite{van2008tsne}, where different color represents different sound categories. Note that multi-label frames are excluded for better display effects.}
  \label{fig:vis}
\end{figure}

\begin{figure}[htp]
  \centering
  \centerline{\includegraphics[scale=0.475]{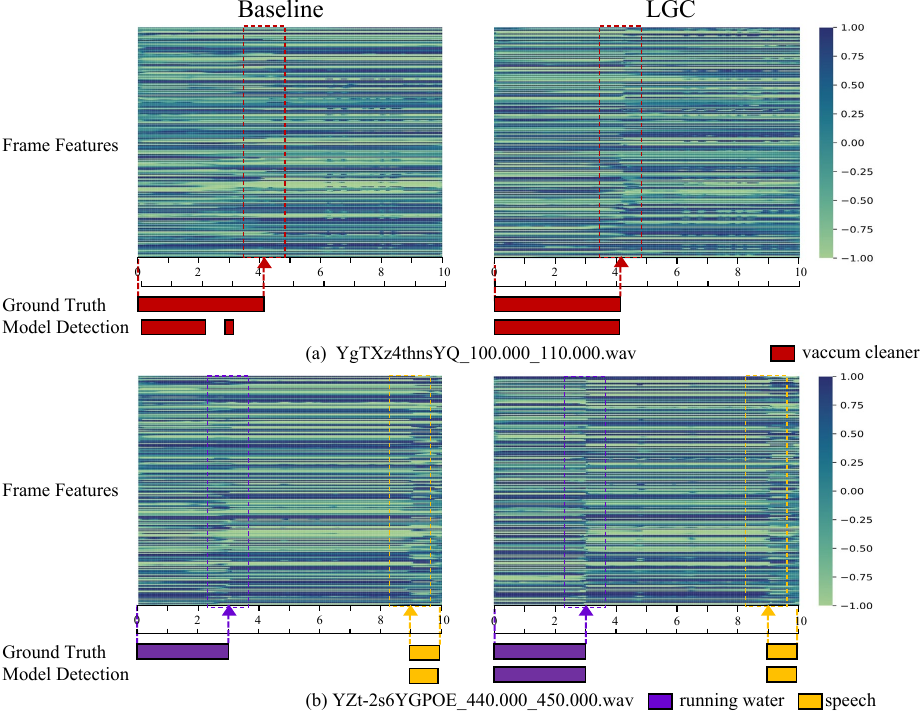}}
  \vspace{-0.75em}
  \caption{Event predictions (0.50 as the threshold) and frame feature visualizations of Baseline and LGC for audios in the validation set.}
  \label{fig:edge_detection}
\end{figure}

\begin{table}[th]
  \caption{Ablation studies for LGC without a certain component, where LC, GC, SAS, and MP are short for Local Consistency, Global Consistency, Selective Anchor Sampling, and Multiple Prototypes.}
  \setlength{\tabcolsep}{6mm}
  \label{tab:table_three}
  \centering
  \begin{tabular}{cccc}
    \toprule
    \multicolumn{1}{c}{\textbf{w/o}} & {\textbf{EB-F1}(\%)} & {\textbf{PSDS$_{1}$}} & {\textbf{PSDS$_{2}$}} \\
    \midrule
    - & \bf 48.8 & \bf 0.392 & \bf 0.579 \\
    LC & 45.6 & 0.374 & 0.575  \\
    GC & 44.5 & 0.373 & 0.570  \\
    SAS & 46.2 & 0.378 & 0.567  \\
    MP & 47.6 & 0.383 & 0.571             \\
    \bottomrule
  \end{tabular}
\end{table}
We evaluate the performance of LGC trained without a specific component, and the results can be found in Table~\ref{tab:table_three}. As illustrated in Table~\ref{tab:table_three}, two consistency regularization methods both contribute to the performance gain, without which an absolute drop of 3.5\% on EB-F1 and 0.02 on PSDS$_{1}$ can be witnessed. We argue that without LC, the model can not learn robust frame representations, while without GC, no global information is available for feature learning. In addition, SAS and MP are also essential to LGC as they assist the student encoder in modeling class-specific features and aligning with prototypes. We further visualize the frame-wise latent representations of a well-trained model for each class and calculate the intra-class variance $S_w$ and inter-class variance $S_b$, the larger ${tr(S_b)}/{tr(S_w)}$ is, the more well-structured feature space that the model learns. As shown in Figure~\ref{fig:vis}, our method LGC enables better intra-class compactness and inter-class dispersion of the feature space compared to the baseline model. Figure~\ref{fig:edge_detection} gives two examples in which LGC makes accurate detections, but the baseline does not. As marked in the figure, the frame-wise features for LGC are more consistent within each sound event while varying dramatically on the boundaries, making it easier for the classifier to discriminate different events.
\vspace{-1.25em}
\vspace{-1.5em}
\section{Conclusion}
In this paper, we develop a novel consistency regularization scheme for semi-supervised SED tasks, which pursues the local consistency on label-level predictions and the global consistency on feature-level representations with the help of audio CutMix and prototypical contrastive learning. Experiments indicate that the proposed method yields superior performance compared to its counterparts and affords promising improvements while complementing existing methods. 

\vfill\pagebreak

\bibliographystyle{IEEEbib}
\bibliography{strings,refs}

\begin{thebibliography}{10}

\bibitem{mesaros2021sound}
Annamaria Mesaros, Toni Heittola, Tuomas Virtanen, and Mark~D Plumbley,
\newblock ``Sound event detection: A tutorial,''
\newblock {\em IEEE Signal Processing Magazine}, vol. 38, no. 5, pp. 67--83,
  2021.

\bibitem{li2023astsed}
Kang Li, Yan Song, Li-Rong Dai, Ian McLoughlin, Xin Fang, and Lin Liu,
\newblock ``Ast-sed: An effective sound event detection method based on audio
  spectrogram transformer,''
\newblock in {\em ICASSP}, 2023.

\bibitem{xu2023pretrain}
Liang Xu, Lizhong Wang, Sijun Bi, Hanyue Liu, and Jing Wang,
\newblock ``Semi-supervised sound event detection with pre-trained model,''
\newblock in {\em ICASSP}, 2023.

\bibitem{tarvainen2017mean}
Antti Tarvainen and Harri Valpola,
\newblock ``Mean teachers are better role models: Weight-averaged consistency
  targets improve semi-supervised deep learning results,''
\newblock in {\em NeurIPS}, 2017, pp. 1195--1204.

\bibitem{xiao2023cmt}
Shengchang Xiao, Xueshuai Zhang, and Pengyuan Zhang,
\newblock ``Multi-dimensional frequency dynamic convolution with confident mean
  teacher for sound event detection,''
\newblock in {\em ICASSP}, 2023.

\bibitem{verma2019ict}
Vikas Verma, Alex Lamb, Juho Kannala, Yoshua Bengio, and David Lopez-Paz,
\newblock ``Interpolation consistency training for semi-supervised learning,''
\newblock in {\em IJCAI}, 2019, pp. 3635--3641.

\bibitem{zhangmixup}
Hongyi Zhang, Moustapha Cisse, Yann~N Dauphin, and David Lopez-Paz,
\newblock ``mixup: Beyond empirical risk minimization,''
\newblock in {\em ICLR}, 2018.

\bibitem{park19e_interspeech}
Daniel~S. Park, William Chan, Yu~Zhang, Chung-Cheng Chiu, Barret Zoph, Ekin~D.
  Cubuk, and Quoc~V. Le,
\newblock ``{SpecAugment: A Simple Data Augmentation Method for Automatic
  Speech Recognition},''
\newblock in {\em INTERSPEECH}, 2019, pp. 2613--2617.

\bibitem{nam2022filteraugment}
Hyeonuk Nam, Seonghu Kim, and Yong-Hwa Park,
\newblock ``Filteraugment: An acoustic environmental data augmentation
  method,''
\newblock in {\em ICASSP}, 2022, pp. 4308--4312.

\bibitem{chapelle2009semi}
Olivier Chapelle, Bernhard Scholkopf, and Alexander Zien,
\newblock ``Semi-supervised learning,''
\newblock {\em IEEE Transactions on Neural Networks}, vol. 20, no. 3, pp.
  542--542, 2009.

\bibitem{koh2021sct}
Chih-Yuan Koh, You-Siang Chen, Yi-Wen Liu, and Mingsian~R Bai,
\newblock ``Sound event detection by consistency training and pseudo-labeling
  with feature-pyramid convolutional recurrent neural networks,''
\newblock in {\em ICASSP}, 2021, pp. 376--380.

\bibitem{shao22_interspeech}
Nian Shao, Erfan Loweimi, and Xiaofei Li,
\newblock ``{RCT: Random consistency training for semi-supervised sound event
  detection},''
\newblock in {\em INTERSPEECH}, 2022, pp. 1541--1545.

\bibitem{kothinti2022tcl}
Sandeep Kothinti and Mounya Elhilali,
\newblock ``Temporal contrastive-loss for audio event detection,''
\newblock in {\em ICASSP}, 2022, pp. 326--330.

\bibitem{park22cmpr}
Sangwook Park, Sandeep~Reddy Kothinti, and Mounya Elhilali,
\newblock ``{Temporal coding with magnitude-phase regularization for sound
  event detection},''
\newblock in {\em INTERSPEECH}, 2022, pp. 1536--1540.

\bibitem{yun2019cutmix}
Sangdoo Yun, Dongyoon Han, Seong~Joon Oh, Sanghyuk Chun, Junsuk Choe, and
  Youngjoon Yoo,
\newblock ``Cutmix: Regularization strategy to train strong classifiers with
  localizable features,''
\newblock in {\em CVPR}, 2019, pp. 6023--6032.

\bibitem{chapelle2005semi}
Olivier Chapelle and Alexander Zien,
\newblock ``Semi-supervised classification by low density separation,''
\newblock in {\em International workshop on artificial intelligence and
  statistics}, 2005, pp. 57--64.

\bibitem{cakir2017crnn}
Emre Cak{\i}r, Giambattista Parascandolo, Toni Heittola, Heikki Huttunen, and
  Tuomas Virtanen,
\newblock ``Convolutional recurrent neural networks for polyphonic sound event
  detection,''
\newblock {\em IEEE/ACM Transactions on Audio, Speech, and Language
  Processing}, vol. 25, no. 6, pp. 1291--1303, 2017.

\bibitem{hu22d_MGA}
Ying Hu, Xiujuan Zhu, Yunlong Li, Hao Huang, and Liang He,
\newblock ``{A Multi-grained based Attention Network for Semi-supervised Sound
  Event Detection},''
\newblock in {\em INTERSPEECH}, 2022, pp. 1531--1535.

\bibitem{snell2017prototypical}
Jake Snell, Kevin Swersky, and Richard Zemel,
\newblock ``Prototypical networks for few-shot learning,''
\newblock in {\em NeurIPS}, 2017, pp. 4080--4090.

\bibitem{saeed2021contrastiveaudio}
Aaqib Saeed, David Grangier, and Neil Zeghidour,
\newblock ``Contrastive learning of general-purpose audio representations,''
\newblock in {\em ICASSP}, 2021, pp. 3875--3879.

\bibitem{niizumi2022byola}
Daisuke Niizumi, Daiki Takeuchi, Yasunori Ohishi, Noboru Harada, and Kunio
  Kashino,
\newblock ``Byol for audio: Exploring pre-trained general-purpose audio
  representations,''
\newblock {\em IEEE/ACM Transactions on Audio, Speech, and Language
  Processing}, vol. 31, pp. 137--151, 2022.

\bibitem{chen2020infonce}
Ting Chen, Simon Kornblith, Mohammad Norouzi, and Geoffrey Hinton,
\newblock ``A simple framework for contrastive learning of visual
  representations,''
\newblock in {\em ICML}, 2020, pp. 1597--1607.

\bibitem{turpault2019desed}
Nicolas Turpault, Romain Serizel, Ankit~Parag Shah, and Justin Salamon,
\newblock ``Sound event detection in domestic environments with weakly labeled
  data and soundscape synthesis,''
\newblock in {\em Workshop on Detection and Classification of Acoustic Scenes
  and Events}, 2019.

\bibitem{mesaros2016ebf1}
Annamaria Mesaros, Toni Heittola, and Tuomas Virtanen,
\newblock ``Metrics for polyphonic sound event detection,''
\newblock {\em Applied Sciences}, vol. 6, no. 6, pp. 162, 2016.

\bibitem{bilen2020psds}
{\c{C}}a{\u{g}}da{\c{s}} Bilen, Giacomo Ferroni, Francesco Tuveri, Juan
  Azcarreta, and Sacha Krstulovi{\'c},
\newblock ``A framework for the robust evaluation of sound event detection,''
\newblock in {\em ICASSP}, 2020, pp. 61--65.

\bibitem{nam22_fdc}
Hyeonuk Nam, Seong-Hu Kim, Byeong-Yun Ko, and Yong-Hwa Park,
\newblock ``{Frequency Dynamic Convolution: Frequency-Adaptive Pattern
  Recognition for Sound Event Detection},''
\newblock in {\em INTERSPEECH}, 2022, pp. 2763--2767.

\bibitem{park2022tbfl}
Sangwook Park and Mounya Elhilali,
\newblock ``Time-balanced focal loss for audio event detection,''
\newblock in {\em ICASSP}, 2022, pp. 311--315.

\bibitem{van2008tsne}
Laurens Van~der Maaten and Geoffrey Hinton,
\newblock ``Visualizing data using t-sne.,''
\newblock {\em Journal of machine learning research}, vol. 9, no. 11, 2008.

\end{thebibliography}

\end{document}